# SEARCH FOR SOURCES OF PRIMARY COSMIC RAYS AT ENERGIES ABOVE 0.1 P EV AT TIEN SHAN

G. Benkó[1], G. Erdös[1], S.I. Nikolsky[2], N.M. Nesterova[2], V.A. Romakhin[2]
A.P. Chubenko[2], A.L. Shchepetov[2], A. Somogyi[1], A. Varga[1]

1. *KFKI Research Institute for Particles and Nuclear Physics of Hungarian Academy*
2. *P.N. Lebedev Physical Institute (FIAN) of Russian Academy of Sciences*

**ABSTRACT**

Analysis of the experimental data of joint Russian - Hungarian (FIAN and KFKI) research was carried out at Tien Shan station on the search of sources of primary cosmic rays (PCR) at energies $E_0 = 10^{14}\text{-}10^{15}$ eV. Maps of extensive air shower (EAS) arrival directions were presented where the standard deviation of mean event numbers in equatorial co-ordinates exceeded a definite value. The excess of EAS number was observed from an extended galaxy object - the SN remnant and some other directions. The young pulsar PSR 0656+14 located near the centre of this object.

**INTRODUCTION** The joint Russian - Hungarian experiment was carried out at P.N. Lebedev Physical Institute (FIAN) Tien-Shan station ($43.04^0$ N, $76.93^0$ E, P=690 г см$^{-2}$). The separate system was used to investigate the PCR anisotropy. This system contained two parts: the CHRONOTRON to determine EAS inclinations and the KLARA to record the EAS in various intervals of the detected particles in different ranges of the time. Electronics [1] of this system had been developed in KFKI Research Institute for Particles and Nuclear Physics of Hungarian Academy of Sciences and was mounted at Tien-Shan station where co-workers of FIAN and KFKI made measurements during more then four years. The total statistics was about 40 millions of EAS. The system detected EAS with $N_e$ (the total number of electrons) from $3 \cdot 10^4$ to $5 \cdot 10^5$ at Tien Shan level ($N_e$ is quantity connected with PCR energy). The main complex FIAN EAS array [2, 3] operated at the same time for the study of different EAS components.

This paper is based on experimental results for search of arrival directions from possible PCR local sources. It was supposed that candidates were directions, where primary neutral particles (gamma-quanta) were not deviate by magnetic fields of Galaxies. Our statistics allows to observed the excess from those particles on the isotropic background of charged particles, if the fraction of neutrals is not less then $\approx 3 \cdot 10^{-4}$ (now we processed $\approx 2 \cdot 10^7$ events). Directions were chosen when the σ - standard relative deviation of mean event numbers exceeded a definite value.

**METHODS OF DETECTION AND PROCESSING** Registration of EAS arrival directions (zenith and azimuth angles in the local system) was carried out by means of nanosecond CHRONOTRON apparatus [4]. It recorded the time difference of the EAS front arrival in 2 m$^2$ scintillation detectors located at the each point of four stations at 20 m from the centre of the array

[2]. Triggering system selected a fourfold coincidence when one or more particles were registered in each station. The registration of EAS number within definite time intervals and an EAS selection by the number of particles was made by means of the KLARA apparatus. Groups of 36 special high stability GM-counters (the area of each counter was 320 cm$^2$) were placed at the same stations. Numbers of EAS particles (electrons) were recorded in eight intervals. The EAS number in every 15 min. and 2 hour's interval was recorded. Signals of the termination for every 5 min., 15 min., hour and day were recorded too.

Distributions of zenith ($\theta$) and azimuth ($\varphi$) angles were obtained. They corresponded to a $\cos^6\theta$ law for $\theta$ and an isotropy for $\varphi$ at $\theta>15^0$.

After that the simulation was initiated to obtain distributions of EAS total electron number $N_e$ and mean value of $N_e$ for every of eight intervals where KLARA detected particles.

The transformation from the local to an equatorial system: Dec. (declination) and RA (right ascension) was done. Angle intervals were selected where the EAS number exceeded a mean value by a few standard deviations in (Dec. × RA) = ($10^0 \times 10^0$) cell. Arrival directions were selected in ranges: Dec. = $(-5 \div +85)^0$ and RA = $(0 \div 360)^0$.

The total number of EAS events and the total time of observation were calculated for every cells, for the cause if all zenith angles within the cell in 15 min. interval had been observed in the interval $\theta = (15-42)^0$, where the accuracy of $\theta$ and $\varphi$ were equal to $3.5^0$.

**PROCESSING RESULTS AND DISCUSSION** The distributions in Dec. and RA were obtained for four $N_e$ intervals with mean values $<N_e>$: 1) Log$<N_e>$ = 4.6, 2) Log$<N_e>$ = 4.9, 3) Log$<N_e>$ = 5.2, 4) Log$<N_e>$ = 5.5 for EAS in the PCR energy range $E_0=10^{14}-10^{15}$. Cells of arrival directions were selected where $\sigma$ (the relative error) = (1.75–2.00) and $\sigma > 2.1$ are shown for first and second $N_e$ intervals in Fig.1, as well for third and fourth $N_e$ intervals in Fig.2.

A few number of cells with the deviation from the mean number exceeded expected "normal" low. It is possible they are directions to cosmic rays sources.

We compared these results with former TIEN-SHAN [5] and KASCADE [6] data, where arrival directions of EAS without muons and hadrons were selected at the same PCR energy $E_0 > (3 - 5) 10^{14}$ eV. The directions for six out of seven EAS from [5] and for 18 out of 53 from [6] (for $\sigma >2.1$) coincided with our selected EAS arrival directions within of errors. The probability of the accidental coincidence is less than $5 \cdot 10^{-3}$ for both sets of events.

There is a probability for those locations of some PCR possible point sources (Crab, Gemins, Cyg. X-3, X-1, Her X-1, X-2 and some nearest Active Galaxies) coincide with arrival directions where we observed the excess, but the accuracy of our results does not prove it definitely.

However the extended (diameter is $25^0$) Galaxy object - the SN remnant exists. It is observed as source of soft X-rays Monogem Ring [7]. The young pulsar PSR 0656+14 located near its centre. It is possible, that pulsar formed from the SN-explosion [8]. We recorded the excess (σ >2.1) of EAS in several $N_e$ intervals in the region of this object. The theoretical interpretation, that the object is possible EAS sources at high energies, was done in [9].

The excess was also observed in the direction of Galaxy disk at region of RA ≅ (270–310)$^0$ и Dec. ≅ (0–45)$^0$ where a lot of radio and X-ray pulsars are concentrated in comparison with other regions [10, 11]. This region is shown by dotted lines in Fig. 1 and 2.

We plan to process all our experimental data with doubled statistics using smaller Dec. and RA cells.

The Russian version of text was published in Izv RAN, ser., fiz., V 69, n 11, p. 1599-1601. (2004). (E-mail: nester@x4u.lebedev.ru Nesterova)

**ACKNOWLEDGEMENTS** Authors thank colleagues of KFKI and FIAN Tien Shan station for the development and exploitation of Tien-Shan array. Authors thank also colleagues of Institute for Nuclear Research and Nuclear Energy of Bulgarian Academy of Sciences who provided G.M counters. We thank A.D Erlykin,, V.P. Pavlyuchenko and A.V. Uryson for help and advice in the process of a work.

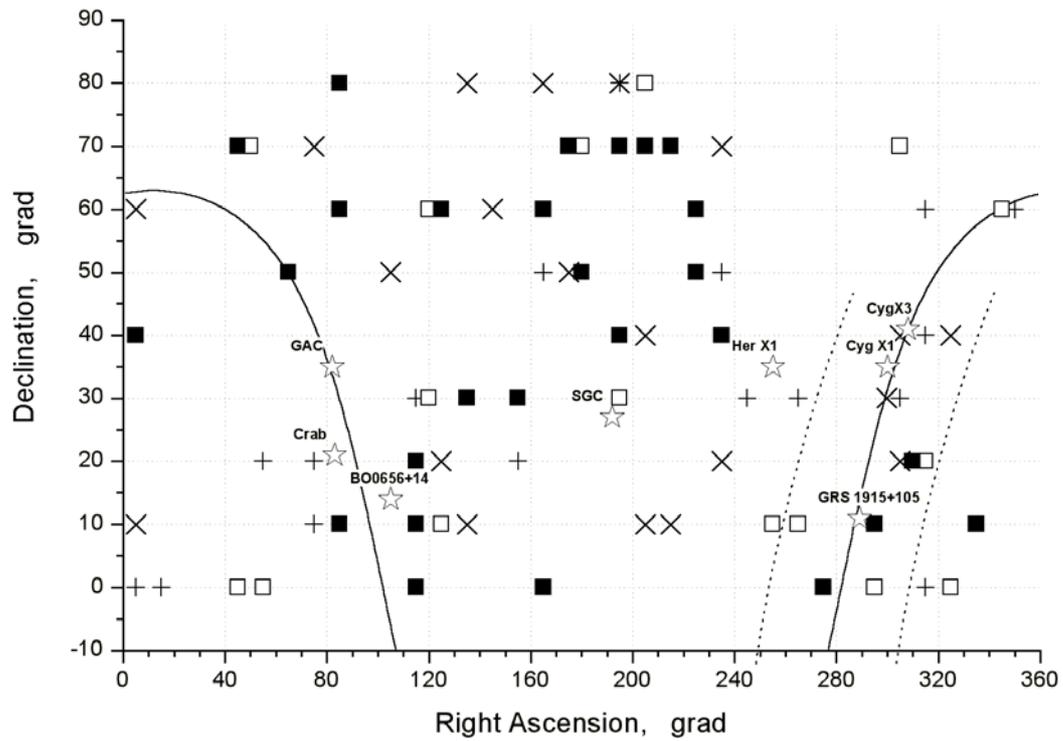

Рис.1 .EAS arrival direction for two Ne-groups. Equatorial coordinates.
Log <$N_e$>=4.6    Log <$N_e$>=4.9

|   |   |   |
|---|---|---|
| × | + | (1.75-2.0) of the standard deviation. |
| □ | ■ | (>+2.1) of the standard deviation. |

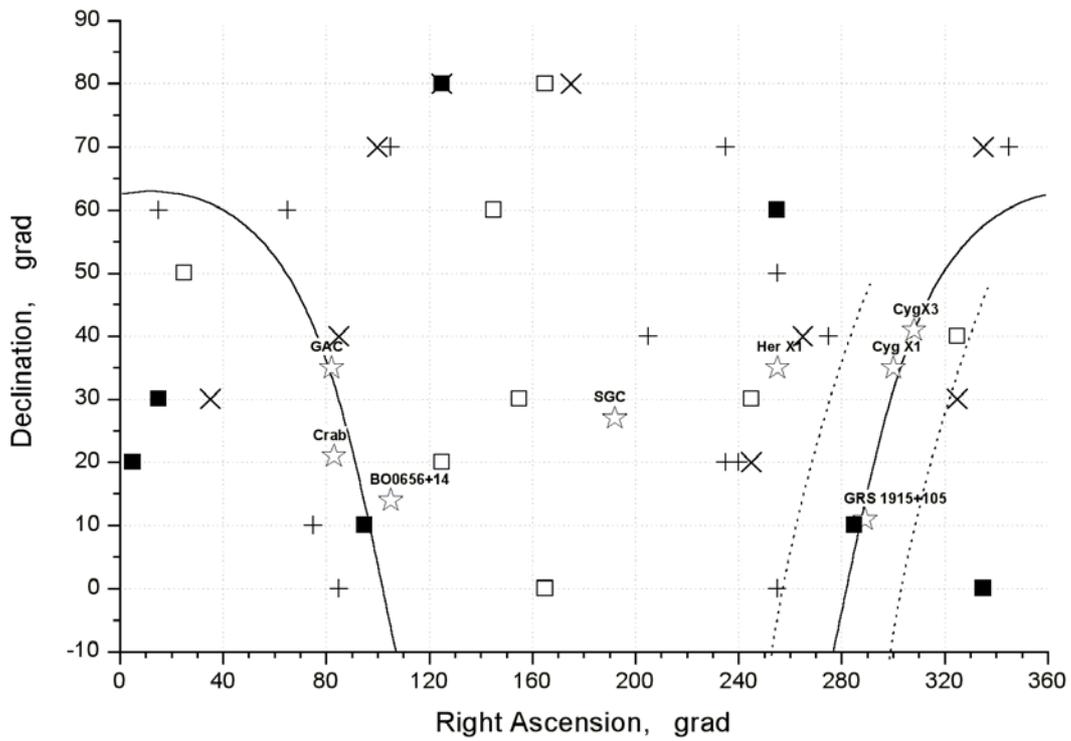

Рис.2 .EAS arrival direction. for two Ne-groups. Equatorial coordinates.
Log <$N_e$>=5.2    Log <$N_e$>=5.5

|   |   |   |
|---|---|---|
| × | + | (1.75-2.0) of the standard deviation. |
| □ | ■ | (>+2.1) of the standard deviation. |